\documentstyle[12pt]{article}
\makeatletter
\@addtoreset{equation}{section}
\makeatother

\def\Lp{\displaystyle{\biggl(}}
\def\Rp{\displaystyle{\biggr)}}
\def\LP{\displaystyle{\Biggl(}}
\def\RP{\displaystyle{\Biggr)}}
\newcommand{\lp}{\left(}\newcommand{\rp}{\right)}

\newcommand{\G}{\Gamma}
\newcommand{\D}{\Delta}

\renewcommand{\d}{\delta}

\newcommand{\f}{\phi}

\newcommand{\m}{\mu}
\newcommand{\n}{\nu}

 \renewcommand{\O}{\Omega}

\newcommand{\s}{\sigma} \renewcommand{\S}{\Sigma}

\renewcommand{\AA}{{\cal A}}
\newcommand{\BB}{{\cal B}}
\newcommand{\CC}{{\cal C}}
\newcommand{\DD}{{\cal D}}

\newcommand{\LL}{{\cal L}}

\newcommand{\NN}{{\cal N}}

\newcommand{\QQ}{{\cal Q}}
\newcommand{\RR}{{\cal R}}
\newcommand{\SS}{{\cal S}}

\newcommand{\complex}{{\kern .1em {\raise .47ex
\hbox {$\scriptscriptstyle |$}}
    \kern -.4em {\rm C}}}
\newcommand{\real}{{{\rm I} \kern -.19em {\rm R}}}
\newcommand{\rational}{{\kern .1em {\raise .47ex
\hbox{$\scripscriptstyle |$}}
    \kern -.35em {\rm Q}}}
\renewcommand{\natural}{{\vrule height 1.6ex width
.05em depth 0ex \kern -.35em {\rm N}}}

\newcommand{\cb}{{\bar c}}
\newcommand{\half}{\frac 1 2}
\newcommand{\pa}{\partial}
\newcommand{\pad}[2]{{\frac{\partial #1}{\partial #2}}}
\newcommand{\fud}[2]  {{\displaystyle{\frac{\delta #1}{\delta #2}}}}

\newcommand{\eg}{{\em e.g.\ }}

\newcommand{\sla}{\raise.15ex\hbox{$/$}\kern -.57em}

\newcommand{\twiddle}{\lower.9ex\rlap{$\kern -.1em\scriptstyle\sim$}}
\newcommand{\afv}{{{\rm all\ fields\ }\varphi}}

\newcommand{\vf}{{\varphi}}

\renewcommand{\=}{&=&} 
\newcommand{\equ}[1]{(\ref{#1})}

\newcommand{\eq}{\begin{equation}}
\newcommand{\eqn}[1]{\label{#1}\end{equation}}
\newcommand{\eea}{\end{eqnarray}}
\newcommand{\eqa}{\begin{eqnarray}}
\newcommand{\eqan}[1]{\label{#1}\end{eqnarray}}
\newcommand{\ba}{\begin{array}}
\newcommand{\ea}{\end{array}}
\newcommand{\eqac}{\begin{equation}\begin{array}{rcl}}
\newcommand{\eqacn}[1]{\end{array}\label{#1}\end{equation}}

\renewcommand{\pad}[2]{{\displaystyle{\frac{\partial #1}{\partial #2}}}}

\newcommand{\hr}{\hat r}
\newcommand{\hBB}{\hat \BB}
\newcommand{\intx}{\int d^4 \! x \, }

\newcommand{\es}{\\[2mm]}

\newcommand{\journal}[4]{{\em #1~}#2\,(19#3)\,#4;}
\newcommand{\aihp}{\journal {Ann. Inst. Henri Poincar\'e}}
\newcommand{\hpa}{\journal {Helv. Phys. Acta}}

\newcommand{\ijmp}{\journal {Int. J. Mod. Phys.}}

\newcommand{\pr}{\journal {Phys. Rev.}}
\newcommand{\jetpl}{\journal {JETP Lett.}}
\newcommand{\prl}{\journal {Phys. Rev. Lett.}}

\newcommand{\cmp}{\journal {Comm. Math. Phys.}}

\newcommand{\zp}{\journal {Z. Phys.}}
\newcommand{\np}{\journal {Nucl. Phys.}}
\newcommand{\pl}{\journal {Phys. Lett.}}

\newcommand{\prep}{\journal {Phys. Reports}}

\newcommand{\nc}{\journal {Nuovo Cim.}}

\newcommand{\annp}{\journal {Ann. Phys. (N.Y.)}}

\makeatletter
\@addtoreset{equation}{section}
\makeatother

\advance\hoffset by -0.6cm
\begin{document}
\centerline{ {\LARGE {\sc
            {\bf UNIVERSIT\'E DE GEN\`EVE}}}}
\centerline{ \raisebox{0mm}{{\small SCHOLA GEVENENSIS MDLIX}} }
\vspace{72mm}

\begin{figure}
\includegraphics{bigsigle.ps}
\end{figure}
\centerline{\LARGE Adler-Bardeen theorem and vanishing }  \vspace{2mm}

\centerline{\LARGE of the gauge beta function }
\vspace{9mm}

\centerline{O. Piguet$^1$\footnotetext[1]{Supported in part
                          by the Swiss National Science Foundation.}
 and  S.P. Sorella$^{1}$}
\centerline{{\small D\'epartement de Physique Th\'eorique}}
\centerline{{\small 24, quai Ernest Ansermet}}
\centerline{{\small CH -- 1211 Gen\`eve 4 (Switzerland)}}

\vspace{10mm}

\centerline{{\normalsize {\bf UGVA---DPT 1992/07--774}} }
\vspace{2cm}

\centerline{\Large{\bf Abstract}}\vspace{4mm}

\noindent The proof of the non-renormalization theorem for the gauge anomaly
of four-dimensional theories is extended to the case of models with a
vanishing one-loop gauge beta function.
\setcounter{page}{0}
\thispagestyle{empty}

\vfill\pagebreak
\section{ Introduction }    \label{section1} 
The non-renormalization
theorem~\cite{adler-bardeen,bardeen,zee,low-schr,padua,bbbc,ps}
 of the four-dimensional
gauge an\-om\-aly~\cite{abbj} is of fundamental
importance for the construction of consistent high energy physics
models. This theorem states that the anomaly coefficient  vanishes at
all orders of perturbation theory if it vanishes in the one-loop
approximation.

The original proof of Bardeen~\cite{bardeen} of the theorem for the
non-Abelian
gauge anomaly is based on an analysis of Feynman graphs: one shows that
if the one-loop triangle anomaly cancels, then there exists
a gauge invariant regularization valid to all orders. Later on it was
recognized by Zee~\cite{zee}
in the Abelian case, and by Costa et al.~\cite{padua} in the non-Abelian
case, that it is possible to give a proof based on the combined use of
the gauge (or BRS) Ward identities and of the Callan-Symanzik equation.
In the same time Lowenstein and Schroer~\cite{low-schr}, and later on
Bandelloni et al.~\cite{bbbc}, achieved, with the quantum action
principle~\cite{lam} as the main tool, an algebraic,
regularization independent, version of the previous proof.

The main advantage of a regularization independent proof is that it can
be naturally  extended to more sophisticated theories,
\eg supersymmetric gauge theories and topological theories,
for which no regularization preserving all the symmetries
is available.

The regularization independent proofs given up to the present
time~\cite{low-schr,bbbc,ps}, as well as the proof given
in~\cite{padua}, based on dimensional regularization,
although very general, have their
domain of validity restricted by the ''technical''
assumption that the one-loop
beta function for the gauge coupling~\cite{beta1}
should not identically vanish.
Even if this assumption covers a very large class of models including
the standard model, there is a wide set of interesting theories for
which the one-loop gauge beta function do indeed vanish. This set
includes in particular gauge models with $N=1$ supersymmetry which may
have some relevance in the construction of grand unified
theories~\cite{ibanez}. Moreover the supersymmetric gauge models with a
vanishing one-loop gauge beta function~\cite{parkes,hamidi,rajpoot} are
the starting point towards the construction of ultraviolet finite
theories~\cite{lps}.
It is therefore needed to have a proof
which also applies to the case of a one-loop vanishing gauge beta function.
This is the aim of the present paper.

The demonstration follows the differential geometry
setup of the descent equations which are known to characterize the
anomaly~\cite{zumino,d-violette,brandt}.  It is
the  continuation of a previous work of the authors~\cite{ps}, where a
completely algebraic proof of the non-renormalization theorem was given
in the case of a non-vanishing one-loop gauge beta function.
The main ingredient, as shown in~\cite{ps},
 is the vanishing of the
anomalous dimensions of the differential form operators which are
solutions of the descent equations.

In the proof one has to use the ghost equation shown in~\cite{bps},
which controls the coupling of the Faddeev-Popov ghost $c$. However this
equation
holds only in the Landau gauge, and  we will therefore have to present
our arguments in this particular gauge. The extension of the
non-renormalization theorem to a general linear covariant gauge can be
easily performed by following the techniques of extended BRS
invariance~\cite{ps-extbrs}, as it was done in~\cite{lps-u1}.

\noindent Let us finish this introduction by some remarks.

The proof we are going to present here concerns the
non-supersym\-met\-ric theories for simplicity, the generalization to the
supersymmetric case being apparently straightforward. There is indeed a
supersymmetric version of the descent equations which allows for
an algebraic set up analog to the non-supersymmetric one and which
leads to a unique characterization of the anomaly~\cite{ps-anom,porrati}.

Our proof covers the cases of theories for which the gauge
beta function does not vanish to higher than one-loop order. It does not
hold, as it stands, in cases of
higher order vanishing gauge beta function.

This proof in particular would not apply to the topological theories
which have vanishing beta functions to all orders~\cite{topol}, but to
the present time there is no known exemple of
such a theory having a gauge anomaly, given as a non-trivial solution of the
Wess-Zumino consistency conditions~\cite{stora,topol}.

It is however relevant for the construction of finite supersymmetric
gauge theories~\cite{lps}. Indeed such a construction starts with a
model whose gauge beta function vanishes only in the one-loop
approximation and depends on a certain number of {\em independent}
couplings (a gauge and a few Yukawa couplings).
It is at this stage that the non-renormalization
theorem is needed. The all order vanishing of the whole set of
beta functions is then ensured by requiring the Yukawa couplings
to be a function of the gauge coupling constant, according to
the ''reduction of coupling constants'' theory of
Zimmermann~\cite{reduction}.

\section{ Properties of Yang-Mills theories in the Landau gauge}
\label{section2}  
The purpose of this section is to give a brief summary of the algebraic
properties which characterize a four-dimensional gauge theory quantized in the
Landau gauge~\cite{bps,ps}.

Let us consider a massless gauge theory whose complete classical action
$\S$, using the same notations of ref.~\cite{ps}, reads:

\eq
\S = \S_{\rm inv}\ + \ \S_{\rm gf} \ + \ \S_{\rm ext} \ ,
\eqn{action}

where $\S_{\rm inv}$, $\S_{\rm gf}$ and $\S_{\rm ext}$
are respectively the gauge invariant
action, the Landau gauge fixing term and the external field dependent part.
They are given by:

\eq
\S_{\rm inv} =  \intx \lp - {1\over 4g^2}  F^{a\m\n}F^a_{\m\n}
   + \LL_{\rm matter}(\f,D_\m \f,\lambda_i ) \rp  \ ,
\eqn{actionclass}

\eq
\S_{\rm gf} = \intx{\ }\lp b^a\pa^\m A^a_\m +  \cb^a \pa^\m {(D_\m c)}^a
                       \rp \ ,
\eqn{gaugefix}

\eq
\S_{\rm ext} = \intx \lp - \O^{a\mu} {(D_{\m}c)}^a +
                                  \half \s^a f^{abc}c^bc^c
            -i Y c^a T^a \f \rp \ ,
\eqn{extaction}
where $f^{abc}$ are the structure constant of a simple compact gauge group
$G$, $T^a$ are the generators of the matter representation and
$\{\lambda_i\}$ denote the self coupling constants of the matter fields
$\f$ whose invariant Lagrangian $\LL_{\rm matter}$ is restricted by the usual
power-counting condition.

The invariance of $\S$ under the nilpotent $BRS$
transformations~\cite{brs} (the external fields $\O$, $\s$, $Y$
being kept invariant as usual):

\eq\ba{lcl}
sA^a_\m \= -{(D_\m c)}^a   \ ,\\
sc^a \= {1\over 2} f^{abc}c^b c^c \ ,\\
s\cb^a\= b^a\ , \\
sb^a\=0 \ ,\\
s\f\= -i c^a T^a \f \ ,
\ea\eqn{brs}
is expressed by the classical Slavnov identity:
\eq
\SS(\S) = \intx \LP \fud{\S}{\O^{a\m}}\fud{\S}{A_\m^a}
     + \fud{\S}{\s^a}\fud{\S}{c^a}
     + \fud{\S}{Y}\fud{\S}{\f} +  b^a\fud{\S}{\cb^a} \RP = 0 \ .
\eqn{slavnov}

This identity is assumed to be broken at the quantum level by the gauge
anomaly~\cite{abbj,brsyug}, i.e.:

\eq
\SS(\G) =  \hbar^{n} r \AA {\ }+{\ }O(\hbar^{n+1} ) \ , \qquad
n \ge 2 \ ,
\eqn{anslavnov}
whith
\eq
\AA{\ }={\ }\varepsilon^{\mu\nu\rho\sigma}\intx \pa_{\mu}c^a \Lp
  d^{abc}\pa_{\nu}A^b_\rho A^c_\sigma -
  {\DD^{abcd}\over12}A^b_\nu A^c_\rho A^d_\sigma \Rp   \ ,
\eqn{anomaly}

\eq
\DD^{abcd} = d^{abn}f^{ncd} + d^{acn}f^{ndb} +
             d^{adn}f^{nbc} \ ,
\eqn{DDtensor}
where $d^{abc}$ is the totally symmetric invariant tensor and $\G$ is the
vertex functional

\eq
\G = \S {\ }+{\ }O(\hbar) \ .
\eqn{gammafunct}

One has to note that eq.\equ{anslavnov} implies that the gauge anomaly is
absent at the one loop level, i.e. we consider the case in which the
coefficient $r$ of the one loop triangle diagram
is equal to zero, due to an
appropriate choice of the matter field representation~\cite{abbj}.

In such a situation the Adler-Bardeen theorem~\cite{abbj,padua,bbbc,ps}
states that the coefficient
$r$ in \equ{anslavnov} identically vanishes.

The vertex functional $\G$, besides the anomalous Slavnov identity
\equ{anslavnov}, is known to obey:

\par

i) ${\ }{\ }{\ }$the Landau gauge-fixing condition and the antighost
equation~\cite{pr}
\eq
\fud{\G}{b^a} = \pa A^a   ,\qquad
\fud{\G}{\cb^a} + \pa \fud{\G}{\O^a}  = 0\ ,
\eqn{gaugecond}
\par

ii) ${\ }$the rigid gauge invariance~\cite{pr}

\eq
\RR^a_{\rm rig}\S = \sum_\afv \intx\d^a_{\rm rig}\vf\fud{\S}{\vf}=0\ ,
\eqn{rigid}
\par

iii) the ghost equation~\cite{bps}

\eq
\intx \LP  \fud{\G}{c^a} + f^{abc}\cb^b \fud{\G}{b^c} \RP \ = \D^a  \ ,
\eqn{ghosteq}

\eq
\D^a = \intx \lp f^{abc}\O^{b\m}A^c_\m - f^{abc}\s^b c^c + iYT^a\f \rp\ .
\eqn{ghostbreak}

These conditions, together with \equ{anslavnov}, allow to write a
Callan-Symanzik equation which is Slavnov invariant up to the order
$\hbar^{n}$~\cite{ps}, i.e.:
\eq\ba{rl}
\CC\Gamma{\ }=& {\ }\Lp \mu \pad{\ }{\mu} +
    \hbar \beta_g {\partial {\ }\over \partial g}
   + \hbar \sum_i \beta_i {\partial {\ }\over \partial \lambda_i}
   + \hbar \gamma_A \NN_A
   +\hbar \gamma_{\phi} \NN_{\phi} \Rp \Gamma{\ }                 \es
         =&{\ }\hbar^{n+1} \Delta^{n+1}_c   + O(\hbar^{n+2})   \ ,
\ea\eqn{calla-sym}
where $\m$ denotes the renormalization point, $\D^{n+1}_c$ is an integrated
local polynomial, $(\beta_g,{\ }\beta_i)$ are respectively the beta functions
for the gauge and the self matter couplings and $(\NN_A,{\ }\NN_{\phi})$ are
the Slavnov invariant counting operators:

\eq
\NN_A = \intx \Lp A^{a\mu}\fud{}{A^{a\mu}} - b^a\fud{}{b^a}
        -{\bar c}^a\fud{}{{\bar c}^a} -
          \Omega^{a\mu}\fud{}{\Omega^{a\mu}} \Rp  \ ,
\eqn{counting1}
\par

\eq
\NN_{\phi} = \intx \Lp \phi\fud{}{\phi} - Y\fud{}{Y}   \Rp \ .
\eqn{counting2}

The vanishing up to the order $\hbar^{n}$ of the ghost anomalous dimension,
i.e. the absence in \equ{calla-sym} of the Slavnov invariant counting term

\eq
\NN_c \G = \intx \Lp c^a \fud{\G}{c^a} - \s^a \fud{\G}{\s^a}   \Rp \ ,
\eqn{counting3}
is due to the ghost equation \equ{ghosteq}.

Moreover, as shown in~\cite{ps}, the use of the Landau gauge allows to define
a renormalized anomaly insertion

\eq
[\AA \cdot \G] = \AA{\ }+{\ }O({\hbar \AA}) \ ,
\eqn{aninsert}
which possesses the following properties:

\eq
\CC[\AA \cdot \G] = \hbar\SS_\G [\hat\D\cdot\G]  + O(\hbar^{n+1})\ ,
\eqn{cs-anom}

\eq
\SS_\G [\AA \cdot \G] = O(\hbar^n) \ ,
\eqn{aninvariance}
where $\SS_\G$ is the linearized Slavnov operator

\eq\ba{rl}
\SS_\G =&\intx \LP \fud{\G}{\O^{a\m}}\fud{}{A^a_\m}
     + \fud{\G}{A^a_\m}\fud{}{\O^{a\m}}
     + \fud{\G}{\s^a} \fud{}{c^a}
     + \fud{\G}{c^a}  \fud{}{\s^a} \es
    &{\ }{\ }{\ }{\ }+ \fud{\G}{Y}\fud{}{\f} + \fud{\G}{\f} \fud{}{Y}
     + b^a\fud{}{\cb^a} \RP
\ea\eqn{slavnovlin}
and

\eq
\SS_\G \SS_\G = O({\hbar^n}) \ .
\eqn{nilpotency}
The non-vanishing right-hand-side of the
last equation is due to the presence of the gauge anomaly in the
Slavnov identity \equ{anslavnov}.

Equations \equ{cs-anom}, \equ{aninvariance} tell us that the insertion
$[\AA \cdot \G]$ obeys a Callan-Symanzik equation without anomalous
dimension (up to a $\SS_\G$-variation) till the order $\hbar^n$,
and that it is Slavnov invariant
up to the order $\hbar^n$.

As we will see in the next sections, properties \equ{cs-anom},
\equ{aninvariance} will provide a complete algebraic proof of the
Adler-Bardeen theorem also in the case of vanishing one loop gauge
beta function.

\section{ Order $\hbar^{n+1}$ } \label{section3}  

Following~\cite{bbbc}, we can extend the anomalous Slavnov identity
\equ{anslavnov} to the order $\hbar^{n+1}$ as

\eq
\SS(\Gamma){\ }={\ }r{\hbar}^n [\AA \cdot \G]{\ }+{\ }
 {\hbar}^{n+1}\BB{\ }{\ }+O({\hbar}^{n+2})  \ ,
\eqn{extsl}
where $[\AA \cdot \G]$ is the anomaly insertion defined in equations
\equ{cs-anom}, \equ{aninvariance} and $\BB$ is an integrated local functional
of ultraviolet dimension four and ghost number one.

Applying the Callan-Symanzik operator to both sides of equation \equ{extsl}
and making use of eq.\equ{cs-anom} and of the algebraic property

\eq
\CC\SS(\Gamma){\ }={\ }\SS_{\Gamma}\CC\Gamma  \ ,
\eqn{relalg}
we get, to the lowest order (i.e. order $n+1$) in $\hbar$, the equation

\eq
\Lp \beta_g^{(1)} \pad{r}{g}{\ }
   +{\ }\sum_i \beta_i^{(1)}\pad{r}{\lambda_i} \Rp \AA{\ }+{\ }
   \mu \pad{\BB}{\mu}{\ }={\ }\SS_{\Sigma}( \D^{n+1}_c - r \hat\D )  \ ,
\eqn{nonren1}
where $\SS_\S$ is the linearized nilpotent operator corresponding to the
classical Slavnov identity \equ{slavnov} and
$(\beta^{(1)}_g,{\ }\beta^{(1)}_i)$ are the
one loop beta functions~\cite{beta1}

Taking into account that $\BB$ is homogeneous of degree zero in the mass
parameter~\cite{bbbc}, i.e.:

\eq
\mu\pad{\BB}{\mu} = 0   \ ,
\eqn{BBind}
and that the gauge anomaly $\AA$ cannot be written as a local
$\SS_\S$-variation, it follows that \equ{nonren1} is equivalent to the
two conditions:

\eq
 \beta_g^{(1)} \pad{r}{g}{\ }
   +{\ }\sum_i \beta_i^{(1)}\pad{r}{\lambda_i} {\ }={\ }0  \ ,
\eqn{cond1}
and

\eq
\SS_\S ( \D^{n+1}_c - r \hat\D) {\ }={\ }0 \ .
\eqn{cond2}

In the case in which the one loop gauge beta function $\beta^{(1)}_g$
does not identically vanish, \equ{cond1} implies the Adler-Bardeen
theorem~\cite{padua,bbbc,ps}.
However, for the time being, we keep \equ{cond1} just as an algebraic
equation in view of the fact that we will allow the coefficient
$\beta^{(1)}_g$ to vanish.
In this case eq.\equ{cond1} implies only that $r$ does not depend on the
self matter couplings $\lambda_i$.

Let us turn now to the analysis of the second condition \equ{cond2}.
This equation shows that the difference $(\D^{n+1}_c - r \hat\D)$,
being a Slavnov invariant quantity, can be expanded in terms of the
elements of the invariant basis~\cite{pr}:

\eq
\Lp {\partial \S \over \partial g},
\qquad {\partial \S \over \partial \lambda_i}, \qquad \NN_A\S,
 \qquad \NN_{\phi}\S, \qquad \NN_c\S \Rp \ .
\eqn{set}
This amounts to rewrite the Callan-Symanzik equation \equ{calla-sym} as

\eq
\CC\G{\ }+{\ } \hbar^{n+1}\gamma_c \NN_c \G{\ }={\ }
     r \hbar^{n+1} \hat\D{\ }+{\ }O(\hbar^{n+2}) \ ,
\eqn{callansym2}
where the ghost-anomalous dimension has reappeared, in agreement with the
fact that its absence is ensured only up to the order $\hbar^{n}$~\cite{ps}.

Finally, repeating the same argument as in~\cite{bbbc}, the Callan-Symanzik
equation \equ{callansym2} extends to the order $\hbar^{n+2}$ as

\eq
\CC\G{\ }+{\ } \hbar^{n+1}\gamma_c \NN_c \G{\ }={\ }
     r \hbar^{n+1}[ \hat\D \cdot \G]{\ }+{\ }\hbar^{n+2} \D^{n+2}_c {\ }+
     O(\hbar^{n+3}) \ ,
\eqn{callansym3}
where $\D^{n+2}_c$ is a local integrated functional.
The interesting feature of this equation is that the general local polynomial
$\D^{n+1}_c$ of eq.\equ{calla-sym} has been replaced by the term
$[\hat \D\cdot\G]$,
which
is the same quantity as the one appearing
in the Callan-Symanzik equation for the
anomaly insertion \equ{cs-anom}. This step will turn
out to be very useful in the discussion of the model at
the order $\hbar^{n+2}$.

\section{ Order $\hbar^{n+2}$ } \label{section4}  

This section is devoted to the analysis of the anomalous Slavnov identity
at the order $\hbar^{n+2}$, i.e. to the algebraic characterization of the
local polynomial $\BB$ in eq.\equ{extsl}.

To do this  we will use property \equ{aninvariance} which shows that the
anomaly insertion $[\AA \cdot \G]$ is Slavnov invariant up to the order
$\hbar^n$.
Applying the linearized operator $\SS_\G$ \equ{slavnovlin} to both sides of
equation \equ{extsl} and making use of \equ{aninvariance} and of the
exact relation

\eq
\SS_\G \SS(\G) = 0 \ ,
\eqn{exact}
we find, to the lowest order in $\hbar$ (remember that $n \ge 2$), the
equation

\eq
\SS_\S \BB = 0 \ .
\eqn{bbinv}

This condition implies that the local polynomial $\BB$ is
Slavnov invariant with ghost number one,
and then can be written as

\eq
\BB{\ }={\ }\hr \AA{\ }+{\ }\SS_\S \QQ  \ ,
\eqn{bbexpr}
where $\AA$ is the gauge anomaly \equ{anomaly}, $\hr$ is an arbitrary
coefficient and $\QQ$ a local integrated polynomial of dimension four and
ghost number zero.
Moreover, since $\BB$ appears in the Slavnov equation \equ{extsl} at the
order $\hbar^{n+1}$, it follows that the cohomological trivial term
$\SS_\S \QQ$ can be reabsorbed in the effective action $\G$ as a local
counterterm without affecting properties \equ{cs-anom},
\equ{aninvariance} and the Callan-Symanzik equation \equ{callansym3}.

The Slavnov identity \equ{extsl} becomes then:

\eq
\SS(\G){\ }={\ }r{\hbar}^n [\AA \cdot \G]{\ }+{\ }
 \hr {\hbar}^{n+1}\AA{\ }+O({\hbar}^{n+2})  \ ,
\eqn{extsl1}
and extends to the order $\hbar^{n+2}$ as

\eq
\SS(\G){\ }={\ }r{\hbar}^n [\AA \cdot \G] {\ }+{\ }
 \hr {\hbar}^{n+1}[\AA \cdot \G]{\ }+{\hbar}^{n+2}\hBB{\ }+{\ }
  O(\hbar^{n+3})   \ ,
\eqn{extsl2}
where $\hBB$ is an integrated local polynomial of ultraviolet dimension four
and ghost number one.
It is important to note that we cannot iterate the previous arguments to
characterize $\hBB$, i.e. property \equ{aninvariance} allows to characterize
only the order $\hbar^{n+1}$.

Commuting now the Callan-Symanzik equation \equ{callansym3} with the Slavnov
identity \equ{extsl2} and using eqs. \equ{cs-anom}, \equ{cond1} and the
algebraic relations:

\eq
\NN_c \SS(\G){\ }={\ }\SS_\G \NN_c\G \ ,
\eqn{commut1}

\eq
\NN_c [\AA \cdot \G]{\ }={\ }\AA{\ }+{\ }O(\hbar) \ ,
\eqn{anomcounting}
we get, to the lowest order (i.e. $n+2$) in $\hbar$, the equation

\eq\ba{l}
\Lp
 \beta_g^{(2)} \pad{r}{g}{\ }+{\ }\sum_i \beta_i^{(2)}\pad{r}{\lambda_i}
{\ }+{\ }
 \beta_g^{(1)} \pad{\hr}{g}{\ }+{\ }\sum_i \beta_i^{(1)}\pad{\hr}{\lambda_i}
 \Rp \AA{\ } +{\ }
   \mu \pad{\hBB}{\mu}{\ }\es
={\ }\SS_{\S}( \D^{n+2}_c - \hr \hat\D )  \ ,
\ea\eqn{nonren2}
where $(\beta^{(2)}_g,{\ }\beta^{(2)}_i)$ are the two loop beta functions.

As in the previous section, taking into account that $\hBB$ is homogeneous
of degree zero in the mass parameter and that the anomaly $\AA$ cannot be
written as a local $\SS_\S$-variation, equation \equ{nonren2} splits into
the two equations:

\eq
\SS_{\S}( \D^{n+2}_c - \hr \hat\D ){\ }={\ }0   \ ,
\eqn{dhat}

\eq
\Lp
 \beta_g^{(2)} \pad{r}{g}{\ }+{\ }\sum_i \beta_i^{(2)}\pad{r}{\lambda_i}
{\ }+{\ }
 \beta_g^{(1)} \pad{\hr}{g}{\ }+{\ }\sum_i \beta_i^{(1)}\pad{\hr}{\lambda_i}
 \Rp{\ }={\ }0 \ .
\eqn{twoloop}

Eq.\equ{twoloop}, as it will be discussed in the next chapter, allows to
control the dependence of the anomaly coefficient $r$ from the coupling
constants $(g,{\ }\lambda_i)$ in the case in which $\beta^{(1)}_g = 0$.

As one can easily understand, this is due to the presence in eq.\equ{twoloop}
of the second order beta functions.

\section{ The Adler-Bardeen theorem  } \label{section5}  

As shown in the previous sections, the anomaly coefficient $r$ in
\equ{anslavnov} is constrained by the two conditions \equ{cond1},
\equ{twoloop}; here rewritten for convenience:

\eq
 \beta_g^{(1)} \pad{r}{g}{\ }
   +{\ }\sum_i \beta_i^{(1)}\pad{r}{\lambda_i} {\ }={\ }0  \ ,
\eqn{one}

\eq
\Lp
 \beta_g^{(2)} \pad{r}{g}{\ }+{\ }\sum_i \beta_i^{(2)}\pad{r}{\lambda_i}
{\ }+{\ }
 \beta_g^{(1)} \pad{\hr}{g}{\ }+{\ }\sum_i \beta_i^{(1)}\pad{\hr}{\lambda_i}
 \Rp{\ }={\ }0 \ .
\eqn{two}

To discuss the consequencies of these equations on the coefficients
$(r,{\ }\hr)$ let us consider first the case in which the one loop gauge
beta function $\beta^{(1)}_g$ is nonvanishing.

In this case, as shown in~\cite{padua,bbbc,ps}, eq.\equ{one} implies the
Adler-Bardeen theorem, i.e. that $r=0$.
Equation \equ{two} reduces to:

\eq
 \beta_g^{(1)} \pad{\hr}{g}{\ }+{\ }\sum_i \beta_i^{(1)}\pad{\hr}{\lambda_i}
{\ }={\ }0 \ ,
\eqn{hrequ}
from which it follows that also $\hr$ vanishes; improving then the validity
of the Slavnov identity \equ{extsl2} to all orders of perturbation theory by
induction.

Let us consider now the case in which

\eq
 \beta_g^{(1)}{\ }={\ }0{\ }, \qquad
 \beta_g^{(2)}{\ }\ne {\ }0 \ .
\eqn{new}

Equations \equ{one}, \equ{two} become:

\eq
 \sum_i \beta_i^{(1)}\pad{r}{\lambda_i} {\ }={\ }0  \ ,
\eqn{new1}
and

\eq
 \beta_g^{(2)} \pad{r}{g}{\ }+{\ }\sum_i \beta_i^{(2)}\pad{r}{\lambda_i}
{\ }+{\ } \sum_i \beta_i^{(1)}\pad{\hr}{\lambda_i}
{\ }={\ }0 \ .
\eqn{new2}

Eq.\equ{new1} implies that $r$ is independent from the self
matter couplings $\lambda_i$. It follows then that eq.\equ{new2} reads:

\eq
 \beta_g^{(2)} \pad{r}{g}{\ }+ \sum_i \beta_i^{(1)}\pad{\hr}{\lambda_i}
{\ }={\ }0 \ ,
\eqn{new3}
which is easily seen to imply that $r=0$,
howing to the fact that $r$ depends only on the gauge coupling
$g$ and that the two loop gauge beta function $\beta^{(2)}_g$~\cite{beta2}
is not identically zero for vanishing self matter couplings.

This concludes the proof of the Adler-Bardeen theorem in the case of
vanishing  one loop gauge beta function.


\end{document}